%% file: src/arxiv.tex
\documentclass[journal,twoside]{IEEEtran}

\usepackage[utf8]{inputenc}
\usepackage[T1]{fontenc}
\usepackage{hyperref}
\usepackage{url}
\usepackage{booktabs}
\usepackage{amsfonts}
\usepackage{nicefrac}
\usepackage{microtype}
\usepackage{graphicx}
\graphicspath{{media/}}
\usepackage[table,dvipsnames]{xcolor}
\usepackage{multirow}
\usepackage{amsmath,amssymb}
\usepackage{amsthm}
\usepackage{mathrsfs}
\usepackage{textcomp}
\usepackage{algorithm}
\usepackage{algorithmicx}
\usepackage{algpseudocode}
\usepackage{braket}
\usepackage{listings}
\usepackage{subcaption}
\usepackage{mathtools}

\title{Harnessing DEN models for quantum computing tasks on neutral atom QPUs}

\author{\IEEEauthorblockN{Chiara Vercellino\IEEEauthorrefmark{1}\IEEEauthorrefmark{2}\IEEEauthorrefmark{3},
Giacomo Vitali\IEEEauthorrefmark{1}\IEEEauthorrefmark{2},
Paolo Viviani\IEEEauthorrefmark{1},
Alberto Scionti\IEEEauthorrefmark{1},
Olivier Terzo\IEEEauthorrefmark{1},
Bartolomeo Montrucchio\IEEEauthorrefmark{2}}\\
\IEEEauthorrefmark{1}\textit{Fondazione LINKS}, Torino, Italy \\
\IEEEauthorrefmark{2}\textit{\textit{DAUIN}, Politecnico di Torino}, Torino, Italy\\
\IEEEauthorrefmark{3}\textit{chiara.vercellino@linksfoundation.com}
}

\begin{document}
\maketitle

\begin{abstract}
We present our work on effectively representing unit-disk graphs on the registers of neutral atom quantum machines. Specifically, we aimed to embed graphs corresponding to proteins and cellular antenna networks into unit-disk graphs, ensuring compatibility with the registers of two real QPUs: Orion Alpha by PASQAL and Aquila by QuEra. To address machine-specific constraints, we made adjustments and integrated \textit{Distance Encoder Networks} (\textit{DEN}) from our previous work. Despite these challenges, we successfully embedded up to 76\% of protein-representing graphs for a quantum machine learning classification task on the Aquila QPU, and all subgraphs derived from 90 antenna geographical positions in Turin, Italy, on the Orion Alpha QPU. In the latter case, the graphs represented instances of the graph coloring problem, which we tackled using the hybrid quantum-classical algorithm \textit{BBQ-mIS}. These promising results underscore the effectiveness and versatility of our embedding approach for representing unit-disk graphs on neutral atom quantum computers across diverse applications.
\end{abstract}


\input{src/01_intro}
\input{src/03_methodology.tex}
\input{src/04_results.tex}
\input{src/05_conclusion.tex}

\bibliographystyle{IEEEtran}
\bibliography{biblio.bib}

\end{document}

%% file: src/01_intro.tex
\section{Introduction}

Neutral atom quantum machines are a cutting-edge technology that harnesses the principles of quantum mechanics to perform computational tasks. At their core, these machines utilize individual neutral Rubidium (Rb) atoms as qubits.

One fascinating application of neutral atom quantum machines is their ability to represent natively through their machine Hamiltonian unit-disk graphs (UDG). Unit-disk graphs are a type of geometric graph where vertices represent points in a plane, and edges connect vertices if the corresponding points are within a unit distance of each other\cite{clark1990unit}.\\
Neutral atom quantum machines offer a unique advantage in representing unit-disk graphs due to their inherent propertiest. By positioning the atoms in the quantum machine 2D register and exciting them from the ground state $|g\rangle$ to the excited Rydberg  state $|r \rangle$ by laser pulses, the quantum system evolves according to the Hamiltonian $\mathcal{H}$. Specifically, if we consider a system of $n$ qubits, laser settings characterized by $\Omega$ Rabi frequency and $\delta$ detuning and the position in the register of the $i$-th atom is $p_i$, then the machine Hamiltonian is expressed as in \eqref{eq:Hamiltonian}. The coefficient $C_6$ is related to the energy level of the Rb atoms used to represent the $|r\rangle $ state ($C_6=862690 \times 2 \pi$ MHz $\mu m^6$ for $|r \rangle = |70 S_{1/2}\rangle$ of the $^{87} Rb$ atoms), $\hbar$ is the reduced Plank constant \cite{wurtz2023aquila}.
\begin{align}
    & \mathcal{H}=\frac{\Omega \hbar}{2}\sum_{i=1}^n (|g_i\rangle \langle r_i| + |r_i\rangle \langle g_i|) - \delta \hbar \sum_{i=1}^n \hat{n}_i+\sum_{i<j} V_{ij} \label{eq:Hamiltonian} \\
    & V_{ij}=C_6\frac{\hat{n}_i \hat{n}_j}{|p_i-p_j|^6} \label{eq:interactions}\\
    & \hat{n}_i = |r_i\rangle \langle r_i |
\end{align}

As highlighted by the $V_{ij}$ term \eqref{eq:interactions}, interactions between pairs of qubits decrease with distance. In fact, by specifying values of $\Omega$ and $\delta$, we can characterize the Rydberg radius $r_b = \sqrt[6]{ \frac{C_6}{\hslash \sqrt{\Omega^2 + \delta^2}}}$. According to $r_b$, the qubits can incur into the blockade effect: $r_b$ acts as a threshold distance, atoms closer than $r_b$ cannot be simultaneously in the state $|r \rangle$, thus positioning atoms and setting the laser pulses parameters, it is possible to represent UDGs, whose vertices have the coordinates of the corresponding qubits in the register and the edges are present whenever qubits are closer than $r_b$ \cite{ciampini2015ultracold, picken2018entanglement}.
By encoding the graph's vertices onto individual qubits, neutral atoms machines can efficiently explore and process the graph's connectivity and properties to solve problems of interest, such as combinatorial graph optimization or graphs classification. 


In our previous research \cite{vercellino2022neural, vercellino2023neural}, we introduced a neural-network-based heuristic designed to embed Unit Disk Graphs (UDGs) into a prototype neutral atoms platform. We identified certain necessary conditions for graphs to possess feasible unit-disk embeddings, operating under the assumption that atoms could be freely positioned anywhere within the register, \textit{i.e.}, free-space assumption.


However, upon transitioning to experimentation with two real Quantum Processing Units (QPUs), specifically Orion Alpha by PASQAL and Aquila by QuEra, we encountered additional hardware requirements. Our investigations spanned two distinct applications:
\textit{i)} On Orion Alpha, we implemented the \textit{BBQ-mIS} algorithm outlined in \cite{vercellino2023bbq}. This algorithm represents a hybrid quantum-classical approach aimed at solving graph coloring problems.
\textit{ii)} On Aquila, we conducted tests involving a quantum machine learning algorithm inspired by \cite{henry2021quantum}, which aims to classify proteins as either enzymes or non-enzymes.


For our graph coloring application, our dataset comprises 90 antennas located in the city of Turin, extracted from a publicly available website\footnote{https://opencellid.org}. The graph coloring problems are derived from the optimization of Principal Cell Identifiers (PCIs), where each antenna must be efficiently assigned a PCI to minimize conflicts. Specifically, antennas within overlapping coverage areas should not share the same PCI, and the total number of PCIs must be minimized.
To address this application, we adapted the solutions obtained using the DEN model \cite{vercellino2023neural} to map the coordinates in free-space to a predefined set of fixed positions. These positions were determined by a triangular lattice with a side length of $5 \mu m$.


For graph classification, we embedded samples from the PROTEINS dataset \cite{borgwardt2005protein, dobson2003distinguishing} into the Aquila register. To ensure the feasibility of these embeddings, we introduced an additional constraint. While the DEN model accounted for the unit-disk constraint and determined the minimum/maximum distances among atoms, our experimental setup required the inclusion of a row-spacing constraint.
During the preparation of the register, qubits were mandated to either belong to the same row or possess a y-axis distance greater than or equal to $4 \mu m$. This constraint was crucial for accommodating the physical layout of the register and ensuring the viability of our experimental runs.

%% file: src/03_methodology.tex
\section{Methodology}


Because of the unique characteristics of the Aquila and Orion Alpha machines, we opted to adapt and integrate the DEN model's embedding method in two distinct manners. Employing a nearest neighbor approach, we transformed the free-space solutions into positions arranged according to a triangular-based lattice layout (Orion Alpha's QPU).
For embedding the graphs derived from the PROTEINS dataset for the classification task, we introduced a new component to the \textit{Embedding Loss Function} (\textit{ELF}) \cite{vercellino2023neural}. This addition was crucial for accurately representing the row-spacing constraint, ensuring that our embeddings aligned with the physical constraints of the quantum hardware (Aquila's QPU).

\subsection{UDG embedding on Orion Alpha: \textit{DEN} models and nearest neighbor techniques}


 To embed the graph coloring samples onto Orion Alpha hardware, we employed a novel approach involving the training of a DEN model on each sample. Unlike traditional predictive models, the DEN operates as an optimization solver \cite{vercellino2022neural, vercellino2023neural}. For each new instance of the constrained unit-disk graph (CUDG) problem, the \textit{DEN} learns to generate a feasible embedding by leveraging an initial unfeasible yet easily retrievable solution.

This learning process leverages the \textit{Embedding Loss Function} (\textit{ELF}), which imposes penalties by assigning positive values within the loss function to be minimized. These penalties are applied to address constraint violations, ensuring that the DEN model adjusts and learns to generate embeddings that conform to specific constraints inherent in the Constrained Unit Disk Graph (CUDG) problem.

 \begin{itemize}
     \item Atom pairs must maintain a distance $\geq 5 \mu m$ but $\leq 40\mu m$ to adhere to hardware register constraints. 
     \item For the unit-disk constraint, atom pairs corresponding to adjacent vertices must be separated by a distance $\leq 10.26 \mu m$, while atom pairs corresponding to non-adjacent vertices must have a distance $> 10.26 \mu m$.
 \end{itemize}

As anticipated, the actual Quantum Processing Unit (QPU) we intended to utilize mandates that atom positions, representing the vertices of the UDGs, be selected from a predefined set of fixed positions arranged in a triangular lattice layout. This requirement aims to expedite algorithm execution times, particularly due to the calibration procedure. The available positions in the lattice correspond to the coordinates of optical traps within the register, with 61 traps arranged in a triangular layout with a side length of $5 \mu m$.

Given that the DEN model was not originally designed to determine atom positions in discrete space, and integrating the corresponding integrality constraint into the \textit{ELF} proved infeasible due to non-differentiable points in the loss function, our initial approach was to solve the Mixed Integer Program (MIP) using state-of-the-art solvers such as \textit{Gurobi}\cite{bixby2007gurobi}, \textit{Ipopt}\cite{wachter2006implementation}, and \textit{GLPK}. 
However, this approach led to quadratic or polynomially increasing linear constraints, rendering it impractical to retrieve feasible embedding solutions within a time limit of 5 hours, even for small graph instances comprising 7-10 vertices. Consequently, we sought an alternative solution to map the DEN solutions to the positions of the optical traps.


Our final approach, outlined in Algorithm \ref{alg:orion_alpha}, begins with a DEN solution and proceeds to sort the vertices in descending order of degree. Subsequently, it identifies the closest trap position for each vertex. While this approach does not guarantee the preservation of a unit-disk representation of the graph, it consistently yields hardware-feasible embeddings. The decision to prioritize vertices with higher degrees in the sorting process is motivated by the objective of positioning vertices with more complex connectivity, \textit{i.e.}, those with a greater number of adjacent vertices, at earlier stages of the embedding process.

\begin{algorithm}
\caption{Embedding algorithm on Orion Alpha}\label{alg:orion_alpha}
\begin{algorithmic}
\State $P \gets$ DEN solution for $\mathcal{G}=(\mathcal{V},\mathcal{E})$
\State $T \gets$ trap coordinates in the triangular-based lattice
\State $\mathcal{V}_s \gets sort(\mathcal{V})$ \Comment{sort $\mathcal{V}$ by decreasing degree}
\For{$v \in \mathcal{V}_s$}
\State $p \gets P(v)$ \Comment{(x,y) coordinates of $v$ in $P$}
\State $p' \gets$ nearest neighbor of $p$ in $T$
\State $T \gets T\setminus p'$
\State $P(v) \gets p'$
\EndFor
\end{algorithmic}
\end{algorithm}

\subsection{UDG embedding on Aquila: \textit{ELF} with row spacing constraints}\label{sec:Aquila_UDG}

\begin{figure*}[ht!]
\centering
    \includegraphics[width=\textwidth]{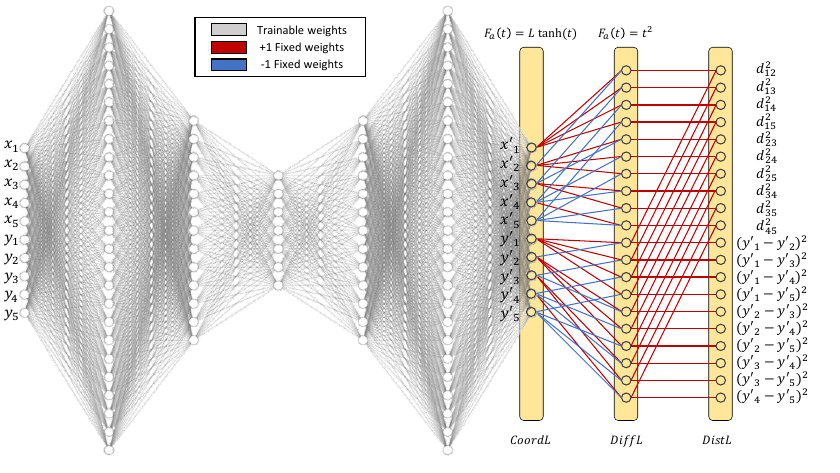}
    \caption{Modified \textit{DEN} model architecture for feasible embeddings on Aquila QPU - an example of embedding a 5-vertex graph.}
    \label{fig:DEN}
\end{figure*}

For UDG embeddings on the Aquila register, we were able to maintain our reliance on free-space assumptions. However, compared to the prototype considered in the definition of the \textit{DEN} model, we encountered two significant differences \cite{wurtz2023aquila}.

Firstly, the Aquila register does not possess a circular shape; instead, it is rectangular, measuring $75 \mu m \times 76 \mu m$. This rectangular shape proved advantageous from the perspective of \textit{ELF} design, as it simplified the modeling of atom placement within the register. By utilizing the activation function in the \textit{CoordL}, we could easily represent atom belonging to the register. Setting the parameter $L=75/2$ enabled us to generate $(x,y)$ coordinates $\in (-75/2, +75/2)^2$. Consequently, by applying a simple translation to the DEN solutions along the $x$ and $y$ axis, we were able to obtain embeddings that could be accommodated within the register.

\begin{figure}[ht!]
\centering
    \includegraphics[width=\linewidth]{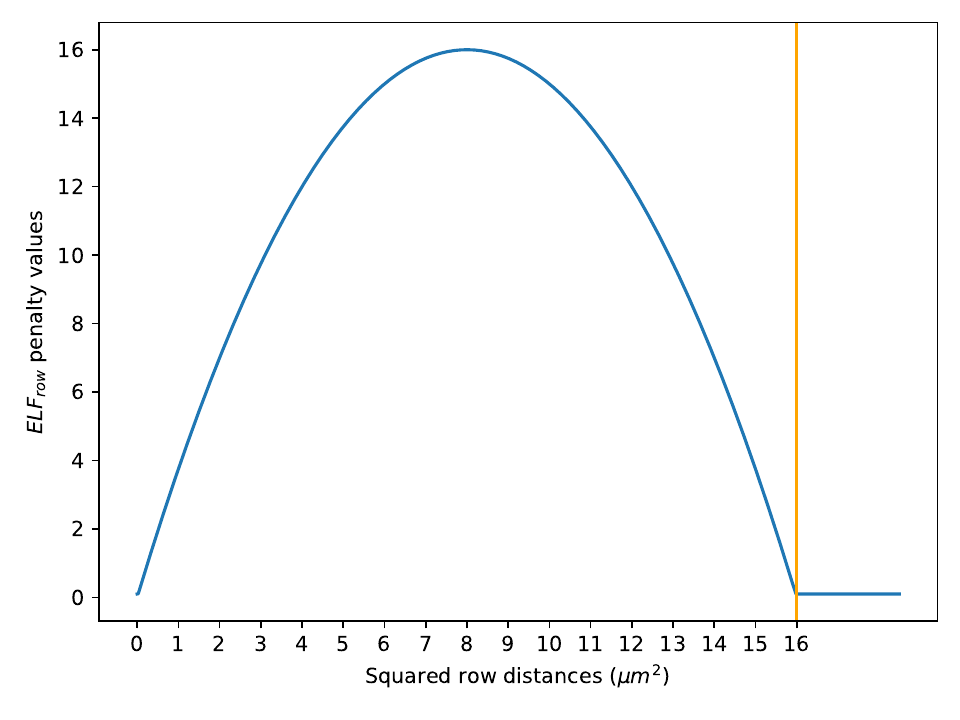}
    \caption{$ELF_{row}$ component for incorporating the row-spacing constraint. $ELF_{row}$ equals 0 in cases where two vertices either share the same row or are separated by a row-distance greater than $4 \mu m$ (squared distance $\geq 16 \mu m^2$).}
    \label{fig:ELF_row}
\end{figure} 

A critical hardware constraint arises from the register loading procedure, significantly impacting the embedding process. To expedite the loading of atoms into the QPU, they are organized into rows within the register. This necessitates that rows be spaced at least $4 \mu m$ apart. In order to accommodate this row-spacing constraint, we made modifications to the DEN architecture, as illustrated in Fig. \ref{fig:DEN}, which depicts the model for a 5-vertex embedding.

Similar to the previous version, the input consists of an infeasible embedding solution represented by the vertex coordinates $x_1, \ldots, x_5, y_1, \ldots, y_5$. These coordinates are passed through an autoencoder comprising fully connected and trainable layers (depicted in gray). At the final layer of the autoencoder, we apply an activation function defined as $F_a(t)=L\cdot tanh(t)$, yielding new coordinates $x'_1, \ldots, x'_5, y'_1, \ldots, y'_5$ that align with the rectangular register.
Subsequent layers possess fixed weights, meaning they are not updated by the \textit{AdamW} optimizer responsible for optimizing the autoencoder weights.

The fixed-weight layers are sparsely populated with values of $\pm 1$. The \textit{DiffL} layer computes the pairwise differences along both the $x$ and $y$ axes, namely $x'_1-x'_2, x'_1-x'_3, \ldots, x'_4-x'_5, y'_1-y'_2, y'_1-y'_3, \ldots, y'_4-y'_5$.
On the \textit{DiffL} layer, we apply the square activation function $F_a(t)=t^2$. The most significant departure from the DEN version presented in \cite{vercellino2023neural} is evident in the output layer, \textit{DistL}. This layer encompasses both the squared pair differences between each vertex pair $i,j$, $d_{ij}^2=(x'_i-x'_j)^2+(y'_i-y'_j)^2$ and the squared row distance $(y'_i-y'_j)^2$.
Considering a graph $\mathcal{G}(\mathcal{V}, \mathcal{E})$, with $\mathcal{V}$ as the set of vertices, $\mathcal{E}$ as the set of undirected edges, and $n=|\mathcal{V}|$, the \textit{DistL} layer consists of $2\times\binom{n}{2}$ elements. The first $\binom{n}{2}$ output elements contribute to the computation of the $ELF_{min}$ and $ELF_{max}$ components of the loss function, as described in \cite{vercellino2023neural}. When the sum of $ELF_{min}$ and $ELF_{max}$ is 0, it indicates that the coordinates $(x'_1, y'_1), \ldots (x'_n, y'_n)$ corresponds to a unit-disk feasible embedding. In this configuration, adjacent vertices are positioned closer than $10.26 \mu m$ and further apart than $4 \mu m$, while non-adjacent vertices are separated by distances exceeding $10.26 \mu m$. Furthermore, they inherently adhere to the register dimensions due to the definition of $F_a$ in the \textit{CoordL}.
To incorporate the row-spacing constraint into the loss function, we introduced a new component, $ELF_{row}$, which operates on the last $\binom{n}{2}$ elements in the \textit{DistL} layer. $ELF_{row}$ combines a parabolic penalty, peaking at a row distance of $2\sqrt{2} \mu m$, with a threshold function to prevent negative contributions from row distances exceeding $4 \mu m$ (refer to Fig. \ref{fig:ELF_row}).

When the sum of $ELF_{min}$, $ELF_{max}$, and $ELF_{row}$ is 0, we achieve a feasible embedding for the Aquila platform.

%% file: src/04_results.tex
\section{Results and Discussion}

To evaluate the effectiveness of these enhanced versions of the DEN model, we conducted experiments involving graph embeddings for two distinct applications. The first application pertained to a combinatorial optimization task, specifically the resolution of the graph coloring problem. The second application focused on a quantum machine learning scenario for graph classification.

In discussing the results of these embeddings, we introduce the following notation: the graph to be embedded is denoted as $\mathcal{G}(\mathcal{V}, \mathcal{E})$, with $n$ vertices. We define $D_{adj}:=\max_{(i,j) \in \mathcal{E}} d_{ij}$ as the maximum distance between adjacent vertices determined by the DEN model, and $d_{\neg adj}:=\min_{(i,j) \notin \mathcal{E}} d_{ij}$ as the minimum distance between non-adjacent vertices. Feasible embeddings are identified by satisfying $D_{adj}\leq 10.26 \mu m$ and $D_{adj}<d_{\neg adj}$. The larger the difference $d_{\neg adj}-D_{adj}$, the higher the quality of the embedding, enabling better discrimination between adjacent and non-adjacent vertices in the presence of noise. Additionally, $\Delta_{max}$ represents the maximum degree in $\mathcal{G}$, while $|K_{max}|$ indicates the size of the maximum estimated fully-connected subgraph in $\mathcal{G}$\footnote{We used \textit{networkx} heuristic to estimate $|K_{max}|$}.

As discussed in \cite{vercellino2022neural}, two necessary conditions to embed a unit-disk graph under the free-space assumption, considering a minimum distance of $4 \mu m$ among vertices and a maximum distance $D_{adj}\leq 10.26 \mu m$, are $\Delta_{max}\leq 18$ and $|K_{max}|\leq 7$.

\subsection{Graph coloring}

\begin{figure*}[ht!]
\centering
    \includegraphics[width=\linewidth]{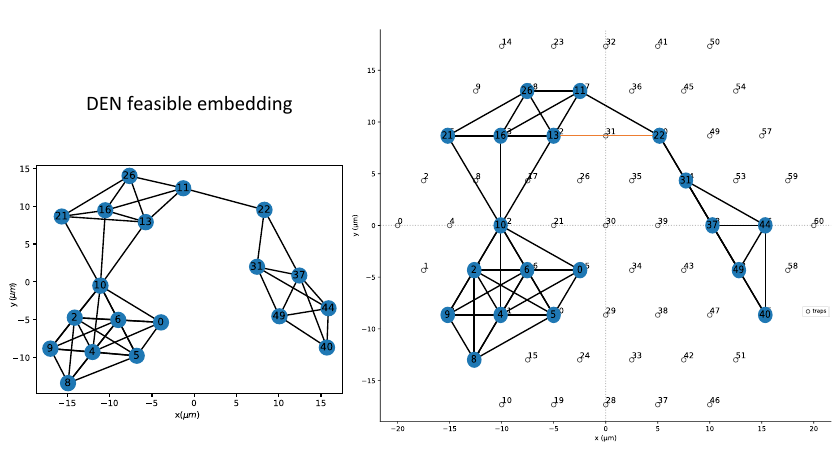}
    \caption{Mapping of the feasible embedding for sample $19G\_1cc$ to Orion Alpha register.}
    \label{fig:orion_emb}
\end{figure*} 

For the graph coloring application, the original dataset represents a cellular antenna network comprising 90 antennas. These networks can naturally be represented as unit-disk graphs, as their conflicts are dependent on distances (in our case, the conflict distance is $140 m$). However, obtaining an Orion Alpha-feasible representation is not achievable by simple rescaling due to the minimum distance requirement among atoms and the fixed qubit positions.

Therefore, we utilize a rescaled embedding as an infeasible solution, serving as input for the \textit{DEN} model. Subsequently, we optimize the trainable weights to attain a feasible solution under the free-space assumption, with $L=41/2$. The maximum distance among Orion Alpha register traps in the lattice is $40 \mu m$, and the minimum distance among atoms is $5 \mu m$. We then apply Algorithm \ref{alg:orion_alpha} to the \textit{DEN} solutions to map coordinates from free-space to the triangular lattice.

Given that the total number of antennas, and thus the number of vertices to be mapped to the QPU register, is 90, exceeding the available 61 atoms, we identify connected components within the antenna network. Each connected component is embedded separately. Upon subdividing the original network into connected components, we obtain:

\begin{itemize}
    \item 9 isolated vertexes
    \item 6 $K_2$, 4 $K_3$, 1 $K_4$, 2 $K_5$ complete graphs, whose unit-disk embeddings are easily identifiable and for which the graph coloring problem has a trivial singleton solution
    \item 2 7-vertex graphs ($7G\_3cc$ and $7G\_13cc$), 1 10-vertex graph ($10G\_4cc$), and 1 19-vertex graph ($19G\_1cc$)
\end{itemize}

Isolated vertices and complete graphs have trivial graph coloring solutions and embeddings. Therefore, our focus was on the embeddings of graphs  $7G\_3cc$, $7G\_13cc$, $10G\_4cc$ and $19G\_1cc$. Table \ref{tab:GC_res} provides a summary of the results and characteristics of these graphs.  $D_{adj}^l$ and $d_{\neg adj}^l$ represent the values of $D_{adj}$ and $d_{\neg adj}$ computed on the lattice coordinates obtained with Algorithm \ref{alg:orion_alpha}. Under the free-space assumption, the DEN model successfully found feasible embeddings for all four graphs. However, the remapping algorithm generally deteriorated the quality of the embedding solution, as $d_{\neg adj}^l-D_{adj}^l$ was consistently lower than $d_{\neg adj}-D_{adj}$. In some cases, this difference was 0, suggesting a need for improvement in the remapping process. Nevertheless, the \textit{BBQ-mIS} algorithm \cite{vercellino2023bbq}, used to solve the corresponding graph coloring problem, proved robust enough to handle these non-ideal unit-disk configurations and find optimal colorings regardless. This resilience can be attributed to the small number of edges that are not properly represented. Figure \ref{fig:orion_emb} illustrates the remapped embedding of graph $19G\_1cc$, where we observe that considering a unit-disk radius $r_b=10.0 \mu m$ would result in only one additional edge (colored in orange) connecting vertices 13 and 22. The issue of $d_{\neg adj}^l-D_{adj}^l=0$ can be resolved by maintaining the left part of the graph's vertex positions as they are and applying the following vertex $\rightarrow$ trap remapppings: 22 $\rightarrow$ 45, 31 $\rightarrow$ 49, 37 $\rightarrow$ 53, 44 $\rightarrow$ 59, 40 $\rightarrow$ 58 and 49 $\rightarrow$ 56 (previously 44 vertex trap).

\begin{table}[htbp]
\caption{Antennas dataset embeddings. All distances are in $\mu m$.}
\begin{center}
\begin{tabular}{|c|c|c|c|c|c|c|}
\hline
Graph & $D_{adj}$ & $d_{\neg adj}$ & $D_{adj}^l$ & $d_{\neg adj}^l$ & $\Delta_{max}$ & $|K_{max}|$\\
\hline
$7G\_3cc$ & 10.2 & 16.3 & 10.0 & 13.2 & 5 & 4\\
$7G\_13cc$ & 10.2 & 12.2 & 10.0 & 10.0 & 5 & 4\\
$10G\_4cc$ & 10.1 & 15.7 & 10.0 & 10.0 & 4 & 4\\
$19G\_1cc$ & 10.2 & 13.3 & 10.0 & 10.0 & 9 & 6 \\
\hline
\end{tabular}
\label{tab:GC_res}
\end{center}
\end{table}

\subsection{Proteins classification}

In the case of the protein classification task, we collected graph samples from the PROTEINS dataset \cite{borgwardt2005protein, dobson2003distinguishing}, and applied the methodology described in Sec. \ref{sec:Aquila_UDG}. Given that the maximum number of atoms, \textit{i.e.}, qubits, in Aquila’s QPU is 256, we specifically selected graph samples with $n\leq256$. Furthermore, as our graph classification application encompassed both emulation of the quantum system on CPU and simulation on QPU, we divided the overall dataset into three subsets:

\begin{itemize}
    \item PROTEINS\_12 includes all the graph samples in the PROTEINS dataset with $n\leq12$
    \item PROTEINS\_16 includes all the graph samples in the PROTEINS dataset with $n\leq16$
    \item PROTEINS\_256 includes all the graph samples in the PROTEINS dataset with $n\leq256$
\end{itemize}

\begin{figure}[ht!]
\centering
    \includegraphics[width=\linewidth]{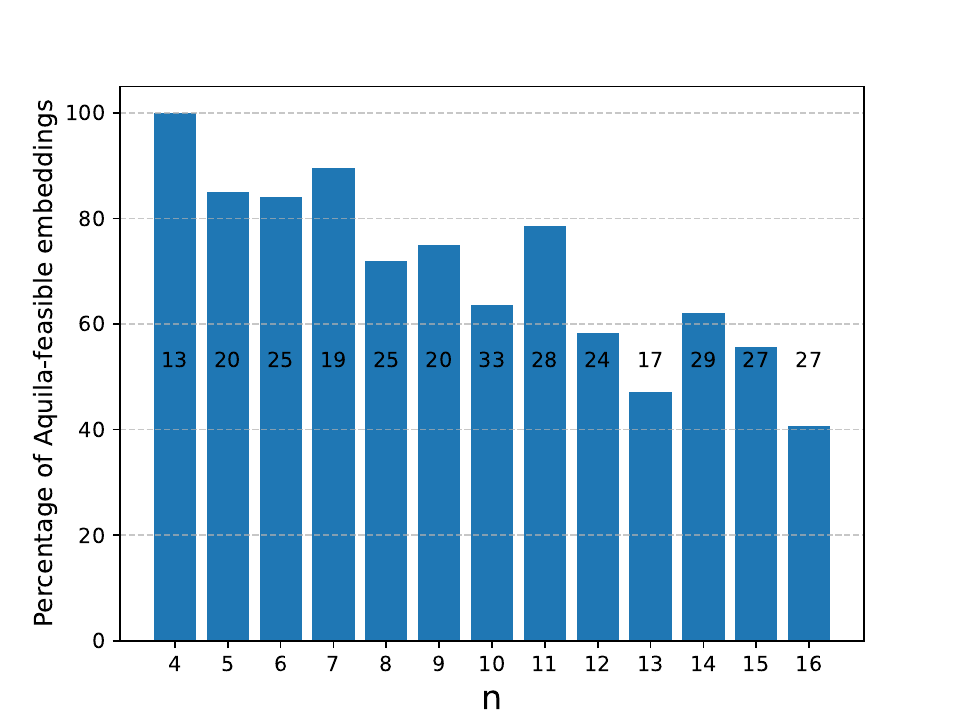}
    \caption{Percentage of graph samples for which the DEN model found an Aquila-feasible embedding, grouped by the number of vertices $n$. The numbers reported in each bin represent the total number of graph samples for each value of $n$.}
    \label{fig:perc_aquila}
\end{figure}

\begin{figure}[ht!]
\centering
    \includegraphics[width=\linewidth]{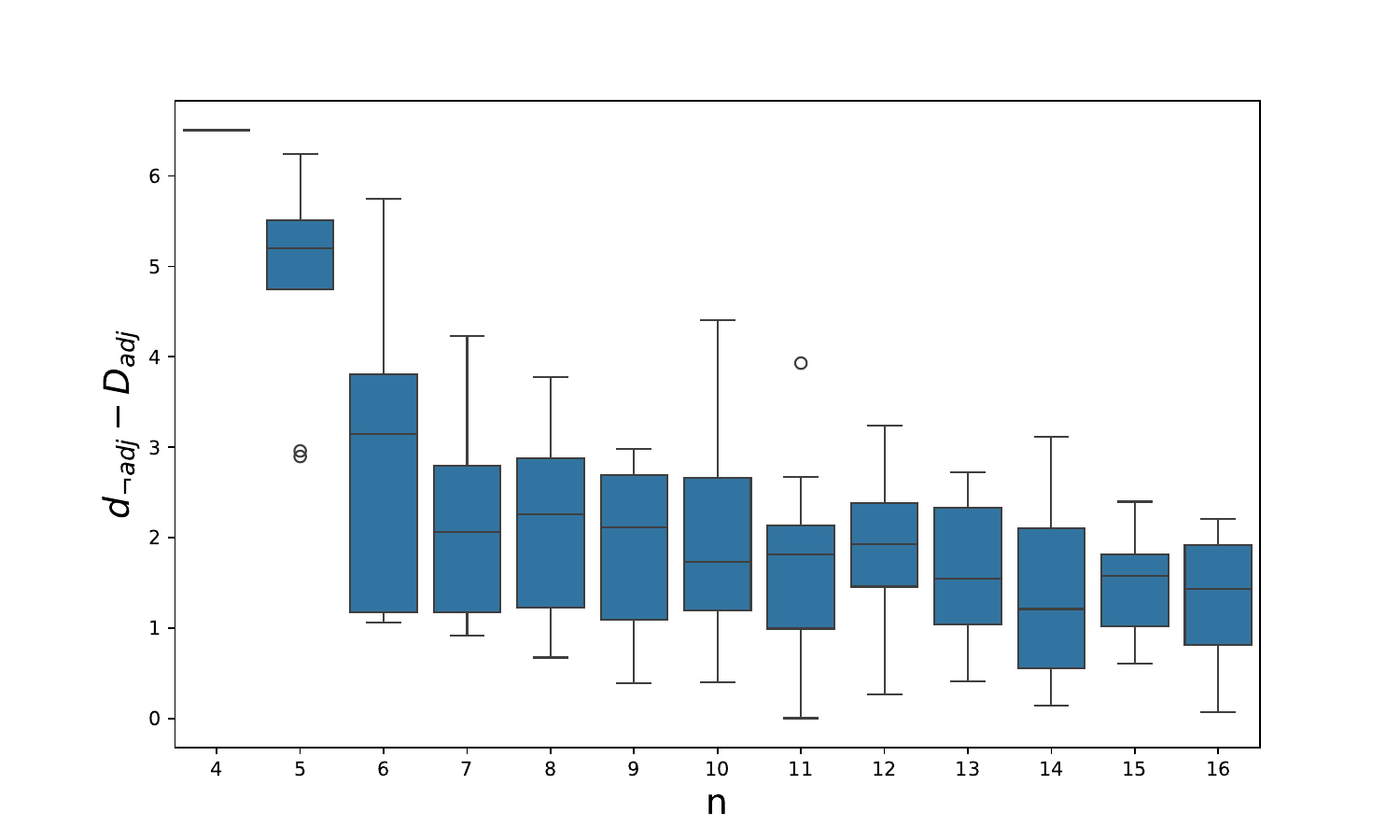}
    \caption{Gap distributions of $d_{\neg adj}-D_{adj}$ grouped by the number of vertices $n$, for the graph-embedded samples in the PROTEINS\_12 and PROTEINS\_16 datasets.}
    \label{fig:gap_aquila}
\end{figure}

We utilized the embedded samples of graphs in PROTEINS\_12 and PROTEINS\_16 for the CPU emulations. The outputs from these emulations were then employed to optimize the laser pulses and identify the best classifier hyperparameters. For both of these datasets, we obtained encouraging results. In the PROTEINS\_12 dataset, we were able to embed 158 out of the total 207 graph samples (approximately 76\% success rate). However, this success rate decreased as we increased the value of $n$. When considering the samples from the PROTEINS\_16 dataset as well, we could obtain Aquila-feasible UDG embeddings for 210 graphs out of 307 samples (approximately 68\% success rate). More detailed results are presented in Fig. \ref{fig:perc_aquila}, which illustrates the percentage of Aquila-feasible embeddings found by the DEN model across different values of $n$. It is noteworthy that as $n$ increases, finding feasible UDG embeddings becomes more challenging, leading to a lower success rate. A similar trend is evident in Fig. \ref{fig:gap_aquila}, which depicts the distributions of the $d_{\neg adj}-D_{adj}$ gap. As the dimensionality of the CUDG problem increases, it becomes increasingly difficult to find embeddings that effectively differentiate between adjacent and non-adjacent vertex pairs.

When we transitioned to the PROTEINS\_256 dataset, obtaining feasible embeddings became even more challenging. We increased the total number of feasible embeddings to 279, which accounts for 34\% of the overall dataset, this increment in dataset size proved beneficial for the classification task. However, the decreased success rate in embedding feasibility highlights the limitation of the current approach, particularly when targeting graphs with $n$ very close to the maximum number of qubits in the QPU.

This outcome is expected, given that positioning 256 vertices within a rectangle measuring $75\mu m \times 76 \mu m$, while adhering to the minimum distance and row-spacing constraints, presents significant difficulties in achieving a feasible exact representation of a general graph.

\begin{figure}[ht!]
\centering
    \includegraphics[width=\linewidth]{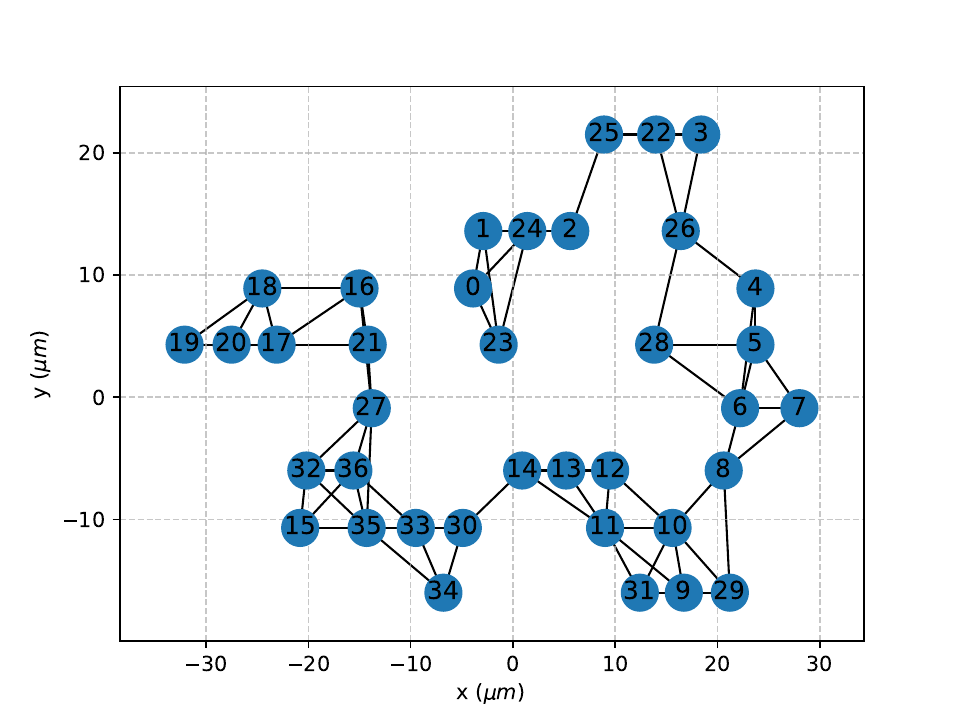}
    \caption{DEN feasible embedding of sample 115 in the PROTEINS\_256 dataset.}
    \label{fig:115_protein_graph}
\end{figure}

To illustrate the complexity of the UDG embedding problem, we present in Fig. \ref{fig:115_protein_graph} the Aquila-feasible embedding of graph sample 115 in the PROTEINS\_256 dataset. This graph comprises 37 vertices, with $\Delta_{max}=7$ and $|K_{max}|=4$. We achieved feasibility with values $D_{adj}=10.0 \mu m$ and $d_{\neg adj}=10.4 \mu m$.

It is noteworthy that, even though the success rate for embedding on larger graphs might seem low, the success rate of state-of-the-art optimization solvers was close to 0\% even on smaller graphs, despite the programming model formulation, and even allowing for very long computational times to find feasible solutions (5-10 hours). This proves the effectiveness of our heuristic, which usually requires less than 5 minutes to provide a solution.

%% file: src/05_conclusion.tex
\section{Conclusion}


In conclusion, this paper thoroughly investigates the embedding of constrained unit-disk graphs (CUDGs) onto neutral atom QPUs, with a focus on the Aquila by QuEra and Orion Alpha by PASQAL platforms. Our experiments and analyses highlight the challenges and constraints in achieving feasible embeddings, particularly as graph complexity and hardware limitations increase.

We address the current lack of tools for neutral atom users by providing a methodology for mapping their problems onto these machines. Despite the difficulty of solving CUDGs, as indicated by the lower success rate of the \textit{DEN} model embedding on larger graphs, our results demonstrate that even partial successes represent progress. Furthermore, we envision that better utilization of available qubits can be achieved with smaller graphs by replicating the same graph multiple times within the register, ensuring sufficient distance between replicates to prevent interaction amongst qubits belonging to different copies. This replication strategy reduces the number of measurements required for quantum tasks, making the process faster and more cost-effective.

